\documentclass[prx,aps,superscriptaddress,nofootinbib,twocolumn]{revtex4-1}

\usepackage{mathrsfs}
\usepackage{amsfonts}
\usepackage{amssymb}
\usepackage{amsmath}
\usepackage{graphicx}
\usepackage[usenames,dvipsnames]{color}
\usepackage[colorlinks=true,citecolor=blue,linkcolor=magenta]{hyperref}
\usepackage{ulem}
\usepackage{lmodern}
\usepackage{amsthm}

\newcommand{\figpath}{.}

\newcommand{\Tr}{\mathrm{Tr}}
\newcommand{\norm}[1]{\Vert #1 \Vert}
\newcommand{\abs}[1]{\vert #1 \vert}

\newcommand{\ket}[1]{\vert{ #1 }\rangle}
\newcommand{\bra}[1]{\langle{ #1 }\vert}
\newcommand{\ketbra}[2]{\vert #1 \rangle \langle #2 \vert}
\newcommand{\braket}[2]{\langle #1 \vert #2 \rangle}

\newcommand{\Dket}[1]{\vert{ #1 }\rangle\rangle}
\newcommand{\Dbra}[1]{\langle\langle{ #1 }\vert}
\newcommand{\Dbraket}[2]{\langle\langle #1 \vert #2 \rangle\rangle}

\newtheorem{theorem}{Theorem}
\newtheorem{lemma}{Lemma}

\begin{document}

\title{Modelling Quantum Devices and the Reconstruction of Physics in Practical Systems}

\author{Hang Ren}
\affiliation{School of Physics, Nankai University, Tianjin 300071, China}

\author{Ying Li}
\email{yli@gscaep.ac.cn}
\affiliation{Graduate School of China Academy of Engineering Physics, Beijing 100193, China}

\begin{abstract}
Modelling quantum devices is to find a model according to quantum theory that can explain the result of experiments in a quantum device. We find that usually we cannot correctly identify the model describing the actual physics of the device regardless of the experimental effort given a limited set of operations. According to sufficient conditions that we find, correctly reconstructing the model requires either a particular set of pure states and projective measurements or a set of evolution operators that can generate all unitary operators.
\end{abstract}

\maketitle

The theory of physics is formalised as a set of definitions and equations that can be applied to all physical systems or a category of systems. In a specific system, physical phenomena are determined by not only the state of the system itself but also physical conditions due to the environment, e.g.~the temperature and external field. Given these conditions, there is a mathematical characterisation, i.e.~a physical model, of the specific system that is consistent with the general theory and the environment. If the general theory is valid, the model is also consistent with experimental results. In quantum theory, the behaviour of a system is stochastic and characterised by the probability distribution according to the corresponding model. An experiment in the quantum system is a sampling of the distribution, and the distribution is directly accessible in the experiment.

\begin{figure}[tbp]
\centering
\includegraphics[width=1\linewidth]{\figpath 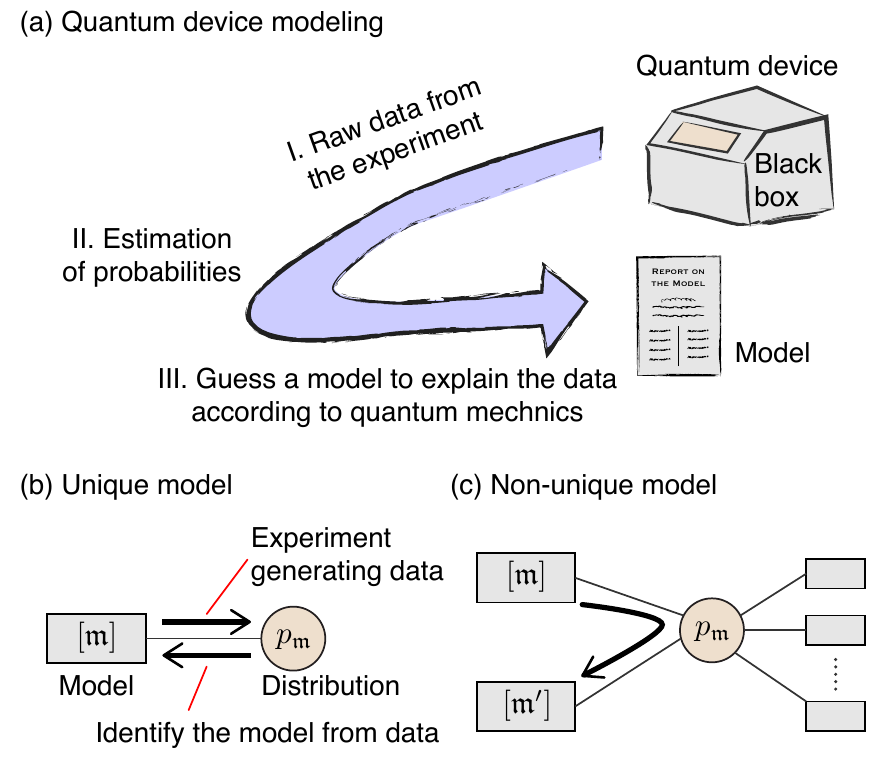}
\caption{
(a) In quantum device modelling, we guess a model that is consistent with the quantum theory and the statistics of data from the experiment. However, such a model may not reflect actual physics in the system. (b) If there is only one model consistent with the corresponding probability distribution, then the model can always be correctly identified. Here, we assume that the statistical fluctuation is negligible provided sufficient data, i.e.~we ignore the error in the estimation of the probability distribution. (c) If there is more than one model consistent with the corresponding probability distribution, the model $[\mathfrak{m}']$ reconstructed from the distribution may be different from the genuine model of the system $[\mathfrak{m}]$ that causes the experimental result.
}
\label{fig:model}
\end{figure}

In this paper, we consider the question of whether it is possible to reconstruct the model of a specific system from experimental results according to quantum theory, see Fig.~\ref{fig:model}(a). In quantum technology, a set of tools have been developed for the verification of a quantum device, e.g.~randomised benchmarking~\cite{Emerson2005, Knill2008} and quantum tomography~\cite{Poyatos1997, Chuang1997, Mayer2018, Sugiyama2018, Thinh2018}. Using the quantum tomography to reconstruct the model of a quantum device, either some prior knowledge of the state preparation or measurement is required, or only partial information can be obtained~\cite{Merkel2013, Greenbaum2015, BlumeKohout2017, DallArno2018}. No matter what technical protocol is used, modelling a quantum system is to seek the inverse map from the distribution back to the physical model. If multiple models lead to the same distribution, the genuine model of the system is not unique and cannot be identified in the experiment. In this paper, we show that given the limited ability to manipulate a system, most models are not unique. We also present two sufficient conditions of the uniqueness. Under these conditions, the model of a system can be reconstructed.

The prior knowledge of the Hilbert space dimension is necessary for the uniqueness. For example, we consider the case that the system is a classical computer pretending to be a qubit, and the computer generates all data in the experiment. If the computer is sufficiently powerful, we are not able to distinguish between the classical computer and a qubit, i.e.~the genuine model of the system cannot be identified. We remark that the knowledge of the dimension is not required in self-testing protocols based on the Bell inequality~\cite{Magniez2005, Sekatski2018}. In general, the final state of a quantum process is $\Tr_{\rm E} [ \mathcal{M}_{\rm SE} (\rho_{\rm SE}) ]$, where $\rho_{\rm SE}$ is the initial state of the system and the environment (SE), and $\mathcal{M}_{\rm SE}$ is the map representing the evolution of SE. In this paper, we focus on the case that the final state of an evolution can always be expressed as $\mathcal{M}_{\rm S} (\rho_{\rm S})$, where $\rho_{\rm S}$ is the initial state of the system, and $\mathcal{M}_{\rm S}$ is a map on the system that is independent of $\rho_{\rm S}$. We can apply results in this paper to SE, but then the knowledge of the Hilbert space dimension of SE is necessary. We remark that if quantum processes are non-Markovian and temporally correlated, we still can obtain a matrix description of quantum processes in the quantum tomography~\cite{Pollock2018, Huo2018}.

In quantum theory, the state of a physical system is described by the wave function, and the evolution is described by the Schr\"{o}dinger equation, which can explain all possible experiments given the correct Hamiltonian. In practice, it is always the case that we can only implement a finite set of experiments in a physical system, i.e.~the ability to manipulate the system is limited. Therefore, we model a practical system as follows. We suppose that the initial state, evolution and observable measured in the final state are controlled by input signals to the system. The initial state is one of states $\{\rho_i\}$. The evolution of the system is one of maps $\{\mathcal{M}_j\}$ or a sequence of these maps. For simplification, we only consider measurement setups with two outcomes true and false, which correspond to positive operator-valued measure(POVM) operators $E_k$ and $\openone - E_k$, respectively. Here $\openone$ is the identity operator. The generalisation to the case of multiple outcomes is straightforward. Then, the system is described by these states, maps and measurement operators, and we call
\begin{eqnarray}
\mathfrak{m} = \{ \{\rho_i\}, \{\mathcal{M}_j\}, \{ E_k \} \}
\end{eqnarray}
a {\it representation} of the model. If the Hilbert space dimension of the system is $d_{\rm H}$, $\{\rho_i\}$ and $\{E_k\}$ are $d_{\rm H}$-dimensional matrices, and $\{\mathcal{M}_j\}$ are linear maps on the space of $d_{\rm H}$-dimensional matrices. The representation is {\it physical}, if $\{\rho_i\}$ are Hermitian, positive semidefinite, and normalised, $\{\mathcal{M}_j\}$ are trace-preserving completely positive maps, and $\{E_k\}$ are Hermitian, positive semidefinite and $E_k \leq \openone$. An experiment is a map from input signals controlling the initial state, evolution and measurement setup to an output signal indicating the measurement outcome. Given input signals labeled by $(i, j_1, \ldots, j_N, k)$, the output signal is true or false, and true occurs with the probability
\begin{eqnarray}
p_\mathfrak{m}(i, j_1, \ldots, j_N, k) = \Tr[ E_k \mathcal{M}_{j_N} \cdots \mathcal{M}_{j_1}(\rho_i) ].
\end{eqnarray}
Here, the evolution is a sequence of $N$ maps.

The probability distribution $p_\mathfrak{m}$ is the information accessible in the experiment. Different representations may result in the same distribution. To show the existence of different but distribution-equivalent representations, we consider the representation
\begin{eqnarray}
\mathcal{T}(\mathfrak{m}) = \{ \{\mathcal{T}^{-1}(\rho_i)\}, \{\mathcal{T}^{-1}\mathcal{M}_j\mathcal{T}\}, \{ \mathcal{T}^*(E_k) \} \},
\end{eqnarray}
where $\mathcal{T}$ is an invertible linear map on the space of matrices, and $\mathcal{T}^*$ is the dual map of $\mathcal{T}$, i.e.~$\Tr[\mathcal{T}^*(A)B] = \Tr[A\mathcal{T}(B)]$ for all matrices $A$ and $B$. Later we will show that a map $\mathcal{T}$ always exists such that both $\mathfrak{m}$ and $\mathcal{T}(\mathfrak{m})$ are physical, e.g.~$\mathcal{T}$ can be a unitary or antiunitary transformation. The difference between two representations $\mathfrak{m}$ and $\mathcal{T}(\mathfrak{m})$ is a similarity transformation on maps. Therefore, probability distributions of two representations are the same, i.e.~
\begin{eqnarray}
p_\mathfrak{m}(i, j_1, \ldots, j_N, k) = p_{\mathcal{T}(\mathfrak{m})}(i, j_1, \ldots, j_N, k)
\end{eqnarray}
for all input signals. Because distributions are the same, we cannot distinguish two representations in the experiment, which is the gauge freedom problem of the gate set tomography~\cite{Merkel2013, Greenbaum2015, BlumeKohout2017}.

Two physical representations are always related by a transformation $\mathcal{T}$ if they contain a complete set of states and measurements and have the same probability distribution (see Sec.~\ref{sec:GST}). However, in general, there could exist two representations that are distribution-equivalent but not related by a transformation $\mathcal{T}$.

Some representations are different but describe the same physics. According to Wigner's theorem~\cite{Wigner1931, Uhlhorn1963, Bargmann1964}, unitary and antiunitary transformations do not change probability distributions in all possible experiments (that are not limited by the ability to manipulate the system) and only change the way that we describe the system. If two representations are related by a unitary or antiunitary transformation, they describe the same physics. We use
\begin{eqnarray}
[\mathfrak{m}] = \{ \mathcal{S}(\mathfrak{m})~ \vert ~ \mathcal{S}\in \mathbb{S} \},
\end{eqnarray}
to denote the set of representations equivalent to $\mathfrak{m}$ up to a unitary or anti-unitary transformation. Here, $\mathbb{S}$ denotes the set of unitary and anti-unitary transformations (see Sec.~\ref{sec:Wigner} for details). We call $[\mathfrak{m}]$ a {\it model}.

We are only interested in physical representations. If the representation $\mathfrak{m}$ is physical, the representation $\mathcal{S}(\mathfrak{m})$ is also physical, where $\mathcal{S}\in \mathbb{S}$. However, in general the representation $\mathcal{T}(\mathfrak{m})$ may not be physical.
\begin{lemma}
Physical gauge transformation. Let $\mathcal{T}$ be an invertible linear map. If $\mathfrak{m}$ and $\mathcal{T}(\mathfrak{m})$ are both physical representations with a complete set of states, then $\mathcal{T}$ is Hermitian- and trace-preserving.
\label{the:gauge}
\end{lemma}
The proof is in Sec.~\ref{sec:Lemma1}. A Hermitian-preserving and trace-preserving linear transformation $\mathcal{T}$ can always be expressed in the form $\mathcal{T}(\bullet) = \sum_l \eta_l F_l \bullet F_l^\dag$, where $\{ \eta_l \}$ are real, and $\sum_l \eta_l F_l^\dag F_l = \openone$. Such transformations are used to describe the evolution of an open system with initial correlations between the system and environment~\cite{Jordan2004}.

Next we will show that, for most physical representations $\mathfrak{m}$, there always exists a transformation $\mathcal{T}$ such that $\mathcal{T}(\mathfrak{m})$ is a physical representation but $[\mathfrak{m}]$ and $[\mathcal{T}(\mathfrak{m})]$ are different models, i.e.~$\mathfrak{m}$ and $\mathcal{T}(\mathfrak{m})$ cannot be related by any unitary or antiunitary transformation. In this case, the model that can explain the corresponding probability distribution is not unique, i.e.~the model cannot be identified in the experiment.

A physical model $[\mathfrak{m}]$ is {\it unique} if it contains all distribution-equivalent physical representations. We use
\begin{eqnarray}
\langle \mathfrak{m} \rangle = \{ \mathfrak{m}' ~ \vert ~ \mathfrak{m}'\text{ is physical and }p_{\mathfrak{m}'} = p_\mathfrak{m} \},
\end{eqnarray}
to denote the set of distribution-equivalent physical representations. If $[\mathfrak{m}] = \langle \mathfrak{m} \rangle$, $[\mathfrak{m}]$ is unique. See Fig.~\ref{fig:model}.

\begin{theorem}
Necessary condition of uniqueness. A nontrivial physical model $[\mathfrak{m}]$ is unique only if either one of the states or one of the Choi matrices is not full rank, i.e.
\begin{eqnarray}
\prod_i \det(\rho_i) \times \prod_j \det(C_j) = 0,
\label{eq:determinant}
\end{eqnarray}
where $C_j$ is the Choi matrix of the map $\mathcal{M}_j$.
\label{the:necessary}
\end{theorem}

A model is trivial if all initial states are the maximally mixed state, all maps are unital, and all measurement operators are proportional to identity, i.e.~
$$
\rho_i = d_{\rm H}^{-1}\openone,~~\mathcal{M}_j(\openone) = \openone,~~E_k = \Tr(E_k)\openone.
$$
If the model is trivial, the state of the system is always the maximally mixed state. In the following, we focus on nontrivial models.

To prove Theorem 1, we introduce the depolarising map in the form
\begin{eqnarray}
\mathcal{D}_F(\rho) = F\rho + (1-F)\Tr(\rho)d_{\rm H}^{-1}\openone,
\label{eq:depolarising}
\end{eqnarray}
where $F \geq 0$. If $0 \leq F \leq 1$, the map is physical, $F$ is the fidelity of the map, and it transforms the state into the maximally mixed state with the probability $1-F$. If $F > 1$, the map is not completely positive but still Hermitian-preserving and trace-preserving. We can find that $\mathcal{D}_F \mathcal{D}_{F'} = \mathcal{D}_{FF'}$, therefore $\mathcal{D}_F^{-1} = \mathcal{D}_{F^{-1}}$ is the inverse map of $\mathcal{D}_F$.

The state $\mathcal{D}_F(\rho_i)$ is physical if $F \leq (1-d_{\rm H}\lambda_i)^{-1}$, where $\lambda_i$ is the minimum eigenvalue of $\rho_i$. According to Eq.~(\ref{eq:depolarising}), the minimum eigenvalue of the output state $\mathcal{D}_F(\rho_i)$ is $F\lambda_i + (1-F)d_{\rm H}^{-1}$, which is nonnegative if and only if the output state is physical. Here, we have used that $\rho_i$ is normalised, i.e.~$\Tr(\rho_i) = 1$.

Given an orthonormal basis of the Hilbert space $\{ \ket{a}~\vert~a = 1,\ldots,d_{\rm H} \}$, the Choi matrix of $\mathcal{M}_j$ is $C_j = \mathcal{I}\otimes\mathcal{M}_j(d_{\rm H}\ketbra{\Phi}{\Phi})$, where $\ket{\Phi} = \frac{1}{\sqrt{d_{\rm H}}}\sum_a \ket{a}\otimes\ket{a}$. Then, the Choi matrix~\cite{Choi1975} of $\mathcal{D}_F\mathcal{M}_j$ is
\begin{eqnarray}
C_{F,j} &=& \mathcal{I}\otimes\mathcal{D}_F\mathcal{M}_j(d_{\rm H}\ketbra{\Phi}{\Phi}) = \mathcal{I}\otimes\mathcal{D}_F(C_j) \notag \\
&=& FC_j + (1-F)d_{\rm H}^{-1}\openone\otimes\openone.
\end{eqnarray}
Here we have used that $\Tr_2(C_j ) = \Tr_2(d_{\rm H}\ketbra{\Phi}{\Phi}) = \openone$ ($\mathcal{M}_j$ is trace preserving), and $\Tr_2$ denotes the partial trace of the second subsystem of the tensor product space. The map $\mathcal{D}_F\mathcal{M}_j$ is physical if its Choi matrix $C_{F,j}$ is positive semidefinite. Because the minimum eigenvalue of $C_{F,j}$ is $F\lambda'_j + (1-F)d_{\rm H}^{-1}$, where $\lambda'_j$ is the minimum eigenvalue of $C_j$, the map $\mathcal{D}_F\mathcal{M}_j$ is physical if $F \leq (1-d_{\rm H}\lambda'_j)^{-1}$. Therefore, the map $\mathcal{D}_F\mathcal{M}_j\mathcal{D}_{F^{-1}}$ is physical when $1 \leq F \leq (1-d_{\rm H}\lambda'_j)^{-1}$.

When $F\geq 1$, $\mathcal{D}_{F^{-1}}$ is a physical map, and $\mathcal{D}_{F^{-1}}^*(E_k)$ is a physical measurement operator.

For a physical representation $\mathfrak{m}$, the representation $\mathcal{D}_{F^{-1}}(\mathfrak{m})$ is also physical if $1 \leq F \leq (1-d_{\rm H}\lambda_{\rm min})^{-1}$, where $\lambda_{\rm min} = \min\{ \lambda_i, \lambda'_j \}$ is the minimum eigenvalue of all initial states and Choi matrices in the model. Two models $[\mathfrak{m}]$ and $[\mathcal{D}_{F^{-1}}(\mathfrak{m})]$ are different if $F>1$, unless $[\mathfrak{m}]$ is trivial. See Sec.~\ref{sec:trivial} for details. Therefore, if a model is unique, we must have $\lambda_{\rm min} = 0$, which is equivalent to Eq.~(\ref{eq:determinant}). Theorem 1 is proved.

The necessary condition stated in Theorem 1 can be approached in the experiment. For example, in the thermal state, the probability of an excited state decreases with the temperature. When the temperature is sufficiently low, the reduced density matrix of the thermal state is approximately a singular matrix. In the following, we present two sufficient conditions of uniqueness. However, in both of them a singular state is not sufficient. Either projective measurements or unitary evolution are required.

Because of the restriction due to Eq.~(\ref{eq:determinant}), representations of unique models can only exist in a subset of the representation space. Does a unique model exist? In the quantum process tomography of a qubit, we usually use $\{ \ketbra{0}{0},\ketbra{1}{1},\ketbra{{\rm x}+}{{\rm x}+},\ketbra{{\rm y}+}{{\rm y}+} \}$ as initial states, where $\ket{{\rm x}+}=\frac{1}{\sqrt{2}}(\ket{0}+\ket{1})$ and $\ket{\rm{y}+}=\frac{1}{\sqrt{2}}(\ket{0}+i\ket{1})$, and the same set of operators as measurement operators. If state preparations and measurements of these four operators have fidelity $1$, the model is unique as we prove next. If initial states and measurement operations are known from prior knowledge, the model can be uniquely reconstructed even if state preparations and measurements are not ideal. We would like to remark that in our assumption we do not have any prior knowledge of initial states and measurement operators.

We will prove the general case of a $d$-dimensional Hilbert space. Given an orthonormal basis of the Hilbert space $\{ \ket{a} \}$, we consider projections $\pi_a = \ketbra{a}{a}$, $\pi^{\rm x}_{a,b} = \ketbra{\psi^{\rm x}_{a,b}}{\psi^{\rm x}_{a,b}}$, $\pi^{\rm y}_{a,b} = \ketbra{\psi^{\rm y}_{a,b}}{\psi^{\rm y}_{a,b}}$, $\pi^{\rm x}_{a,b,c} = \ketbra{\psi^{\rm x}_{a,b,c}}{\psi^{\rm x}_{a,b,c}}$ and $\pi^{\rm y}_{a,b,c} = \ketbra{\psi^{\rm y}_{a,b,c}}{\psi^{\rm y}_{a,b,c}}$. Here, $\ket{\psi^{\rm x}_{a,b}} = \frac{1}{\sqrt{2}}(\ket{a}+\ket{b})$, $\ket{\psi^{\rm y}_{a,b}} = \frac{1}{\sqrt{2}}(\ket{a}+i\ket{b})$, $\ket{\psi^{\rm x}_{a,b,c}} = \frac{1}{\sqrt{3}}(\ket{a}+\ket{b}+\ket{c})$ and $\ket{\psi^{\rm y}_{a,b,c}} = \frac{1}{\sqrt{3}}(\ket{a}+i\ket{b}+i\ket{c})$.

\begin{theorem}
Sufficient condition of uniqueness. The physical model $[\mathfrak{m}]$ is unique if $\{\rho_i\},\{E_k\} \supset \Pi$, where $\Pi = \{ \pi_a, \pi^{\rm x}_{a,b}, \pi^{\rm y}_{a,b}, \pi^{\rm x}_{a,b,c}, \pi^{\rm y}_{a,b,c}, \pi^{\rm y}_{c,b,a} ~\vert~ a<b<c \}$.
\label{the:operators}
\end{theorem}

Because of the completeness, taking $\Pi_{\rm QPT} = \{ \pi_a, \pi^{\rm x}_{a,b}, \pi^{\rm y}_{a,b} \}$ as initial states and measurement operators are sufficient for implementing the quantum process tomography. However, they are insufficient for the uniqueness if $d > 2$, which is why we need projections $\pi^{\rm x}_{a,b,c}, \pi^{\rm y}_{a,b,c}, \pi^{\rm y}_{c,b,a}$. We remark that there are not projections in the form $\pi^{\rm x}_{a,b,c}, \pi^{\rm y}_{a,b,c}, \pi^{\rm y}_{c,b,a}$ when $d = 2$, and in this case projections $\Pi_{\rm QPT}$ are sufficient for the uniqueness.

To prove the uniqueness, we assume that there is a distribution-equivalent physical representation $\mathfrak{m}'$. Because of the completeness, $\mathfrak{m}'$ and $\mathfrak{m}$ are related by a similarity transformation on maps, i.e.~$\mathfrak{m}' = \mathcal{T}(\mathfrak{m})$ (see Sec.~\ref{sec:GST})~\cite{Merkel2013, Greenbaum2015, BlumeKohout2017}. Then, $\mathcal{T}^{-1}(\pi)$ and $\mathcal{T}^*(\pi)$ are the initial state and measurement operator in $\mathfrak{m}'$ respectively corresponding to the initial state and measurement operator of the projection $\pi \in \Pi$ in $\mathfrak{m}$.

Two representations $\mathfrak{m}'$ and $\mathfrak{m}$ result in the same distribution. Using $\Tr[\mathcal{T}^*(A)\mathcal{T}^{-1}(B)] = \Tr(AB)$, where $A, B \in \Pi$, we can prove that initial states and measurement operators corresponding to projections in $\Pi$ are related by a unitary or antiunitary transformation, i.e.~there exists $\mathcal{S} \in \mathbb{S}$ such that $\mathcal{T}^{-1}(\pi) = \mathcal{T}^*(\pi) = \mathcal{S}^{-1}(\pi)  = \mathcal{S}^*(\pi)$ for all $\pi \in \Pi$. For other initial states, maps and measurement operators  in $\mathfrak{m}'$, we can think of using initial states and measurement operators $\mathcal{S}^{-1}(\Pi_{\rm QPT})$ to implement the quantum process tomography to reconstruct them. The quantum process tomography requires the prior knowledge of $\mathcal{S}^{-1}(\Pi_{\rm QPT})$. Given $\Pi_{\rm QPT}$, $\mathcal{S}^{-1}(\Pi_{\rm QPT})$ is known up to the transformation $\mathcal{S}$. Therefore, all initial states, maps and measurement operators reconstructed in the quantum process tomography is up to the transformation $\mathcal{S}$, i.e.~$\mathfrak{m}' = \mathcal{S}(\mathfrak{m})$. The detailed proof is in Sec.~\ref{sec:Theorem2}.

In Theorem~\ref{the:operators}, the pure state preparation and projective measurement are required. Next, we present another sufficient condition, in which the state preparation and measurement can be nonideal, but unitary maps are required. A unitary map is a map in the form $\mathcal{U}(\bullet) = u\bullet u^\dag$, where $u$ is a $d_{\rm H}\times d_{\rm H}$ unitary matrix.

\begin{lemma}
Symmetry transformations on unitary maps. Let $\mathbb{U}$ be the set of unitary maps. Then $\mathcal{T}^{-1}\mathcal{U}\mathcal{T} \in \mathbb{U}$ for all $\mathcal{U} \in \mathbb{U}$ if and only if there exists a unitary or antiunitary transformation $\mathcal{S}\in \mathbb{S}$ such that $\mathcal{S}^{-1}\mathcal{U}\mathcal{S} = \mathcal{T}^{-1}\mathcal{U}\mathcal{T}$ for all $\mathcal{U} \in \mathbb{U}$.
\label{the:unitary}
\end{lemma}

In Ref.~\cite{Wolf2008}, it is proved that a physical map is unitary if and only if its determinant is $1$. Based on this result and Wigner's theorem~\cite{Wigner1931, Uhlhorn1963, Bargmann1964}, we can prove Lemma~\ref{the:unitary} (see Sec.~\ref{sec:Lemma2}), which is the dynamics version of Wigner's theorem.

\begin{theorem}
Sufficient condition of uniqueness. The physical model $[\mathfrak{m}]$ is unique if $\{\mathcal{M}_j\} \supset \mathbb{U}$ and $\prod_i \det(\rho_i) = \prod_k \det(E_k) = 0$, i.e.~all unitary maps can be generated, and at least one initial state and one measurement operator are not full rank.
\label{the:maps}
\end{theorem}

For a distribution-equivalent physical representation $\mathfrak{m}'$, it is related to the representation $\mathfrak{m}$ by a transformation $\mathcal{T}$, i.e.~$\mathfrak{m}' = \mathcal{T}(\mathfrak{m})$. This conclusion relies on the fact that a complete set of states can be generated using unitary maps given a nontrivial state, and it is similar for measurement operators. Because $\{\mathcal{M}_j\} \supset \mathbb{U}$, there is a transformation $\mathcal{S}\in \mathbb{S}$ satisfying $[\mathcal{T}\mathcal{S}^{-1},\mathcal{U}] = 0$ for all $\mathcal{U}\in \mathbb{U}$. Then, $\mathcal{T}\mathcal{S}^{-1}$ must be a depolarising map, i.e.~$\mathcal{T} = \mathcal{D}_F\mathcal{S}$. Suppose $\rho \in \{\rho_i\}$ is singular, $\mathcal{S}^{-1}\mathcal{D}_F^{-1}(\rho)$ is physical if and only if $F\geq 1$. Similarly, suppose $E \in \{E_k\}$ is singular, $\mathcal{S}^{*}\mathcal{D}_F^{*}(E)$ is physical if and only if $F\leq 1$. Therefore, we must have $F = 1$ and $\mathcal{T} = \mathcal{S} \in \mathbb{S}$, i.e.~$\mathfrak{m}$ and $\mathfrak{m}'$ are representations of the same model. Theorem~\ref{the:maps} is proved. See Sec.~\ref{sec:Theorem3} for details.

It is easier to approach the sufficient condition in Theorem~\ref{the:maps} in the experiment compared with Theorem~\ref{the:operators}. For a qubit, i.e.~$d_{\rm H} = 2$, a singular state must be a pure state (which can be a mixed state when $d_{\rm H} > 2$). However, the measurement may not be projective. We consider the measurement $\{ E = \eta\ketbra{1}{1}, \openone-E = \ketbra{0}{0}+(1-\eta)\ketbra{1}{1} \}$, where $0< \eta \leq 1$. The POVM operator $E$ is singular, i.e.~the requirement in Theorem~\ref{the:maps} is satisfied. The measurement only reports reliable outcome for one of two states: when the measured state is $\ket{0}$ the outcome is always $0$; when the measured state is $\ket{1}$ the outcome is 1 with the probability $\eta$. Such a nonideal measurement can be approximately realised in the experiment. In some quantum systems, e.g.~trapped ion or superconducting qubit, state preparation and single-qubit gates have high fidelities, but the fidelity of measurement is significantly lower~\cite{Harty2014, Song2019}. Therefore, measurement fidelity is the limiting factor. For trapped ion, the fluorescence is used to implement the measurement~\cite{Olmschenk2019}: if the number of detected fluorescence photons is higher than a threshold, we take the bright state ($\ket{1}$) as the measurement outcome. By setting a high threshold, we can increase the fidelity of detecting the dark state ($\ket{0}$), such that one of two POVM operators can be close to a singular matrix, maybe at the cost of the overall error. For a superconducting qubit, the dispersive measurement is used to distinguish the ground and excited states of the qubit. Compared with the dissipation into the ground state, it is unlikely that the ground state been excited during the measurement. As a result, the measurement of the ground state $\ket{0}$ has a significantly higher fidelity than the excited state $\ket{1}$~\cite{Song2017}. By optimising the setup of the experiment, we can reduce the uncertainty of the quantum tomography caused by the nonuniqueness of the quantum model.

In this paper, we study the uniqueness of quantum models. If a model is unique, it means that the model and the probability distribution of observables in the experiment are one-to-one correspondence, so that we can identify the model using data from the experiment. However, we find that almost all quantum models of practical systems are not unique. A model is unique only if the system can be prepared in a singular state or one of the Choi matrices is singular. We also find that unique models exist, and two sufficient conditions are found. In one of them, pure states and projective measurements are required. In the other, unitary evolution operators are required, but states and measurements can be nonideal. These conditions can only be approximately satisfied in the experiment. Therefore, methods of estimating the uncertainty of the quantum device modelling and using prior knowledge to reduce the uncertainty are interesting, which is out of the scope of this paper. Because the condition of uniqueness is demanding, it is interesting to develop a quantum model meter, which is a system that can be coupled and decoupled with another system such that the model of the whole system is unique, assuming demanding operations are available in the meter system. Given that unique models are rare, developing quantum technologies independent of the gauge freedom is important, e.g.~quantum error mitigation based on the gate set tomography~\cite{Endo2018} and gauge-invariant measure of the gate error~\cite{Lin2019}.

\begin{acknowledgments}
Y.L. thanks Chao Song, Jingning Zhang and Shuaining Zhang for helpful discussions. This work was supported by National Natural Science Foundation of China (Grant No. 11875050) and NSAF (Grant No. U1730449).
\end{acknowledgments}

\appendix

\setcounter{theorem}{0}
\renewcommand{\thetheorem}{\Alph{section}\arabic{theorem}}
\setcounter{lemma}{0}
\renewcommand{\thelemma}{\Alph{section}\arabic{lemma}}

\section{Wigner's theorem: unitary and anti-unitary transformations}
\label{sec:Wigner}

A transformation $u$ on the Hilbert space is called unitary if $\braket{u(\psi)}{u(\phi)} = \braket{\psi}{\phi}$ for all $\psi$ and $\phi$. Such a transformation is linear, i.e.~$\ket{u(\psi)} = u\ket{\psi}$, where $u$ is a unitary matrix.

A transformation $v$ on the Hilbert space is called anti-unitary if $\braket{v(\psi)}{v(\phi)} = \braket{\phi}{\psi}$ for all $\psi$ and $\phi$. Given an orthonormal basis of the Hilbert space $\{ \ket{a} \}$, we can define the complex conjugate transformation $\ket{\psi} = \sum_a \psi_a \ket{a} \rightarrow \ket{w(\psi)} = \sum_a \psi_a^* \ket{a}$, which is anti-unitary. For any anti-unitary transformation $v$, $u = wv$ is always unitary. Therefore, an anti-unitary transformation can always be written as $v = wu$.

\begin{theorem}
Wigner's theorem (version 1). Let $t$ be a transformation on the Hilbert space, and $\abs{ \braket{t(\psi)}{t(\phi)} } = \abs{ \braket{\psi}{\phi} }$ for all $\psi$ and $\phi$. There exists a unitary or anti-unitary transformation $s$ and a function $\alpha$ such that $\ket{t(\psi)} = \alpha(\psi) \ket{s(\psi)}$ for all $\psi$, where $\alpha(\psi)$ is a phase factor.
\end{theorem}

A ray is a set of vectors in the Hilbert space representing the same physical state, i.e.~$[\psi] = \{ \alpha \ket{\psi} \}$, where $\ket{\psi}$ is a non-zero vector, and $\alpha$ is a non-zero complex number. A ray is a one-dimension subspace of the Hilbert space, and we can use the orthogonal projection onto the subspace $\ketbra{\psi}{\psi}$ to represent the ray $[\psi]$, where $\ket{\psi}$ is normalised. Each transformation on the Hilbert space corresponds to a transformation on the ray space, i.e.~the space of rank-one orthogonal projections. For a transformation $t$ on the Hilbert space, the corresponding transformation on the space of rank-one orthogonal projections is $t(\ketbra{\psi}{\psi}) = \ketbra{t(\psi)}{t(\psi)}$. Then, Wigner's theorem can also be stated as follows.

\begin{theorem}
Wigner's theorem (version 2). Let $t$ be a transformation on the space of rank-one orthogonal projections, and $\Tr[ t(\ketbra{\psi}{\psi}) t(\ketbra{\phi}{\phi}) ] = \abs{ \braket{\psi}{\phi} }^2$ for all $\psi$ and $\phi$. There exists a unitary or anti-unitary transformation $s$ such that $t(\ketbra{\psi}{\psi}) = s(\ketbra{\psi}{\psi})$ for all $\psi$.
\end{theorem}

For the $d_{\rm H}$-dimensional Hilbert space, the set of unitary maps on the space of matrices is $\mathbb{U} = \{ \mathcal{U}(\bullet) = u \bullet u^\dag ~\vert~ u\in U(d_{\rm H}) \}$, where $U(d_{\rm H})$ is the group of $d_{\rm H}\times d_{\rm H}$ unitary matrices. Given an orthonormal basis of the Hilbert space $\{ \ket{a} \}$, the transpose map is $\mathcal{W}(\bullet) = \sum_{a,b} \ketbra{b}{a}\bullet\ketbra{b}{a}$. Then, we define $\mathbb{A} = \mathcal{W} \mathbb{U}$ and $\mathbb{S} = \mathbb{U} \bigcup \mathbb{A}$.

We can express a unitary transformation $u$ on the space of rank-one orthogonal projections as $u(\ketbra{\psi}{\psi}) = \mathcal{U}(\ketbra{\psi}{\psi}) = u\ketbra{\psi}{\psi}u^\dag$. On the space of rank-one orthogonal projections, the complex conjugate transformation is equivalent to the transpose transformation $w(\ketbra{\psi}{\psi}) = \mathcal{W}(\ketbra{\psi}{\psi})$. For any anti-unitary transformation $v = wu$, we always have $v(\ketbra{\psi}{\psi}) = \mathcal{W}\mathcal{U}(\ketbra{\psi}{\psi})$. Therefore, we call $\mathbb{S}$ the set of unitary and anti-unitary transformations. Then, Wigner's theorem can be stated as follows.

\begin{theorem}
Wigner's theorem (version 3). Let $t$ be a transformation on the space of rank-one orthogonal projections, and $\Tr[ t(\ketbra{\psi}{\psi}) t(\ketbra{\phi}{\phi}) ] = \abs{ \braket{\psi}{\phi} }^2$ for all $\psi$ and $\phi$. There exists $\mathcal{S} \in \mathbb{S}$ such that $t(\ketbra{\psi}{\psi}) = \mathcal{S}(\ketbra{\psi}{\psi})$ for all $\psi$.
\label{the:Wigner}
\end{theorem}

\section{Lemma 1: physical gauge transformation}
\label{sec:Lemma1}

Because model representations $\mathfrak{m}$ and $\mathcal{T}(\mathfrak{m})$ are both physical, states $\{ \rho_i \}$ and $\{ \mathcal{T}(\rho_i) \}$ are Hermitian and normalised. The set of states $\{ \rho_i \}$ is complete, then any matrix $A$ can always be decomposed in the form $A = \sum_i A_i \rho_i$, where $\{A_i\}$ are decomposition coefficients. Then $\mathcal{T}(A) = \sum_i A_i \mathcal{T}(\rho_i)$. Because $\mathcal{T}(A)^\dag = \sum_i A_i^* \mathcal{T}(\rho_i) = \mathcal{T}(\sum_i A_i^* \rho_i) = \mathcal{T}(A^\dag)$, $\mathcal{T}$ is Hermitian-preserving. Because $\Tr(A) = \Tr[\mathcal{T}(A)] = \sum_i A_i$, $\mathcal{T}$ is trace-preserving. Lemma~1 is proved.

Given an orthonormal basis of the Hilbert space $\{ \ket{a} \}$, we can express a linear map in the form $\mathcal{T}(\bullet) = \sum_{a,b,c,d} T_{c,d,a,b} \ketbra{c}{a} \bullet \ketbra{b}{d}$. The Choi matrix of the map is $C = \mathcal{I}\otimes\mathcal{T} (\sum_{a,b}\ketbra{a}{b} \otimes \ketbra{a}{b}) = \sum_{a,b,c,d} T_{c,d,a,b} \ketbra{a}{b} \otimes \ketbra{c}{d}$. Because the map is Hermitian-preserving and trace preserving, $T_{c,d,a,b}^* = T_{d,c,b,a}$ and $\delta_{a,b} = \sum_{c,d} T_{c,d,a,b} \delta_{c,d}$. Therefore, the Choi matrix is Hermitian, and $\Tr_2(C) = \openone$, where $\Tr_2$ denotes the partial trace of the second subsystem of the tensor product space.

Eigenvalues and eigenvectors of $C$ are $\{ \eta_l \}$ and $\{ \ket{F_l} = \sum_{a,b} F_{l,b,a} \ket{a}\otimes\ket{b} \}$, then the map can be re-expressed in the form $\mathcal{T}(\bullet) = \sum_{l} \eta_l F_l \bullet F_l^\dag$, where $F_{l} = \sum_{a,b} F_{l,a,b} \ketbra{b}{a}$. Here, we have used that $C = \sum_{l,a,b,c,d} \eta_l F_{l,c,a} F_{l,d,b}^* \ketbra{a}{b} \otimes \ketbra{c}{d}$. Because $C$ is Hermitian, $\{ \eta_l \}$ are real. Because $\Tr_2(C) = \openone$, $\sum_{l} \eta_l F_l^\dag F_l = \openone$.

\section{Trivial model}
\label{sec:trivial}

If an initial state $\rho_i$ is not the maximally mixed state, $\mathcal{D}_F(\rho_i)$ is a state with different eigenvalues. Unitary and anti-unitary transformations do not change eigenvalues of a Hermitian matrix, therefore two model representations $\mathfrak{m}$ and $\mathcal{D}_{F^{-1}}(\mathfrak{m})$ cannot be related by any unitary or anti-unitary transformation. It is similar for maps and measurement operators.

If $\mathcal{M}_j$ is not unital, $I_j = \mathcal{M}_j(\openone) \neq \openone$. Because $\mathcal{M}_j$ is trace-preserving, $\Tr(I_j) = d$. Then eigenvalues of $I_j$ and $I_j' = \mathcal{D}_F \mathcal{M}_j \mathcal{D}_{F^{-1}}(\openone) = \mathcal{D}_F(I_j)$ are different if $F > 1$. If $\mathcal{M}_j$ and $\mathcal{D}_F \mathcal{M}_j \mathcal{D}_{F^{-1}}$ are related by a unitary or anti-unitary transformation, i.e.~$\mathcal{D}_F \mathcal{M}_j \mathcal{D}_{F^{-1}} = \mathcal{S}^{-1} \mathcal{M}_j \mathcal{S}$ and $\mathcal{S}\in \mathbb{S}$, we have $I_j' = \mathcal{S}^{-1}(I_j)$. Because $\mathcal{S}$ is unitary or anti-unitary, eigenvalues of $I_j$ and $\mathcal{S}^{-1}(I_j)$ are the same. Therefore, $\mathcal{M}_j$ and $\mathcal{D}_F \mathcal{M}_j \mathcal{D}_{F^{-1}}$ cannot be related by any unitary or anti-unitary transformation.

\section{Vectorised picture and gate set tomography}

\subsection{Vectorised picture}

Matrices can be vectorised according to the Hilbert-Schmidt inner product. For two matrices $A$ and $B$, the inner product is $\Dbraket{A}{B} = \Tr(A^\dag B)$. In the vectorised picture, we use the $d_{\rm H}^2$-dimensional column vector $\Dket{\rho_i}$, square matrix $M_j$, and row vector $\Dbra{E_k}$ to represent the state $\rho_i$, map $\mathcal{M}_j$ and measurement operator $E_k$, respectively. Then the probability can be re-expressed as
\begin{eqnarray}
p_\mathfrak{m}(i, j_1, \ldots, j_N, k) = \Dbra{E_k} M_{j_N} \cdots M_{j_1} \Dket{\rho_i}.
\end{eqnarray}
In the vectorised picture, the difference between two representations $\mathfrak{m}$ and $\mathcal{T}(\mathfrak{m})$ is a similarity transformation on maps, and
\begin{eqnarray}
\Dket{\mathcal{T}^{-1}(\rho_i)} &=& T^{-1}\Dket{\rho_i}, \\
\mathcal{T}^{-1}\mathcal{M}_j\mathcal{T} &=& T^{-1} M_j T, \\
\Dbra{\mathcal{T}^*(E_k)} &=& \Dbra{E_k}T,
\end{eqnarray}
where $T$ is the matrix of $\mathcal{T}$ in the vectorised picture.

The vectorised picture depends on the basis of the matrix space that we choose. Given an orthonormal basis of the Hilbert space $\{ \ket{a} \}$, we take $\{ \ketbra{a}{b} \}$ as the basis of the matrix space. We can find that such a basis is orthonormal. We use $\Dket{a,b}$ to denote the vector of $\ketbra{a}{b}$ in the vectorised picture, then $\Dbraket{c,d}{a,b} = \Tr(\ket{d}\braket{c}{a}\bra{b}) = \delta_{a,c}\delta_{b,d}$. Therefore, we can express the vector as $\Dket{a,b} = \ket{a} \otimes \ket{b}$.

Given the orthonormal basis of the matrix space, a state $\rho$ is represented by the row vector $\Dket{\rho}$ with elements
\begin{eqnarray}
\Dbraket{a,b}{\rho} = \Tr(\ketbra{b}{a}\rho) = \bra{a}\rho\ket{b},
\end{eqnarray}
a linear map $\mathcal{M}$ is represented by the matrix $M$ with elements
\begin{eqnarray}
\Dbra{a,b} M \Dket{c,d} = \Tr[\ketbra{b}{a}\mathcal{M}(\ketbra{c}{d})],
\end{eqnarray}
a measurement operator $E$ is represented by the column vector with elements
\begin{eqnarray}
\Dbraket{E}{a,b} = \Tr(E^\dag\ketbra{a}{b}\rho) = \bra{b}E^\dag\ket{a}.
\end{eqnarray}
If $\mathcal{M}(\bullet) = \sum_q A_q \bullet B_q$ (we can always express a linear map in this form),
\begin{eqnarray}
\Dbra{a,b} M \Dket{c,d} = \sum_q \bra{a}A_q\ketbra{c}{d}B\ket{b}.
\end{eqnarray}
Therefore,
\begin{eqnarray}
\Dket{\rho} &=& \sum_{a,b} \rho_{a,b} \ket{a} \otimes \ket{b}, \\
M &=& \sum_q A_q\otimes B_q^{\rm T}, \\
\Dbra{E} &=& \sum_{a,b} E_{a,b}^* \bra{a} \otimes \bra{b},
\end{eqnarray}
where $\rho_{a,b} = \bra{a}\rho\ket{b}$, ${\rm T}$ denotes matrix transpose, and $E_{a,b} = \bra{a}E\ket{b}$. If $E$ is Hermitian, $E_{a,b}^* = E_{b,a}$.

\subsection{Gate set tomography}
\label{sec:GST}

A model $[\mathfrak{m}]$ is complete if $\{ \rho_i \}$ and $\{ E_k \}$ are both complete sets of Hermitian matrices, i.e.~any $d_{\rm H} \times d_{\rm H}$ Hermitian matrix can always be expressed as a linear combination of $\{ \rho_i \}$ or $\{ E_k \}$. The gate set tomography can be implemented in a complete model.

Suppose $\mathfrak{m}$ and $\mathfrak{m}'$ are two distribution-equivalent physical representations with complete initial states and measurement operators. Because of the completeness, we can find $d_{\rm H}^2$ linearly independent initial states $\{\rho_i \vert i=1,\ldots,d^2_{\rm H}\}$ and $d_{\rm H}^2$ linearly independent measurement operators $\{E_k \vert k=1,\ldots,d^2_{\rm H}\}$ in $\mathfrak{m}$. Corresponding initial states and measurement operators in $\mathfrak{m}'$ are $\{\rho'_i \vert i=1,\ldots,d^2_{\rm H}\}$ and $\{E'_k \vert k=1,\ldots,d^2_{\rm H}\}$, respectively. Then, we have four $d_{\rm H}^2 \times d_{\rm H}^2$ matrices
\begin{eqnarray}
M_{\rm in} &=& \left( \begin{array}{ccc}
\Dket{\rho_1} & \cdots & \Dket{\rho_{d^2_{\rm H}}}
\end{array} \right), \\
M_{\rm out} &=& \left( \begin{array}{c}
\Dbra{E_1} \\ \vdots \\ \Dbra{E_{d^2_{\rm H}}}
\end{array} \right), \\
M'_{\rm in} &=& \left( \begin{array}{ccc}
\Dket{\rho'_1} & \cdots & \Dket{\rho'_{d^2_{\rm H}}}
\end{array} \right), \\
M'_{\rm out} &=& \left( \begin{array}{c}
\Dbra{E'_1} \\ \vdots \\ \Dbra{E'_{d^2_{\rm H}}}
\end{array} \right).
\end{eqnarray}
We have
\begin{eqnarray}
g = M_{\rm out} M_{\rm in} = M'_{\rm out} M'_{\rm in}
\end{eqnarray}
because of the distribution equivalence, where $g_{k,i} = \Tr(E_k \rho_i) = \Tr(E'_k \rho'_i)$.

\begin{lemma}
For two distribution-equivalent physical representations, if one of them is complete, the other is complete.
\end{lemma}

Because of the completeness of $\mathfrak{m}$, $M_{\rm in}$ and $M_{\rm out}$ are both invertible, and $g$ is also invertible. Because $g$ is invertible, $M'_{\rm in}$ and $M'_{\rm out}$ are both invertible, i.e.~$\{\rho'_i \vert i=1,\ldots,d^2_{\rm H}\}$ and $\{E'_k \vert k=1,\ldots,d^2_{\rm H}\}$ are linearly independent complete sets, i.e.~$\mathfrak{m}'$ is complete.

\begin{lemma}
For two distribution-equivalent physical representations $\mathfrak{m}$ and $\mathfrak{m}'$, if they are complete, they are related by a similarity transformation on maps, i.e.~$\mathfrak{m}' = \mathcal{T}(\mathfrak{m})$.
\label{the:completeness}
\end{lemma}

Two representations are related by a transformation $\mathcal{T}$, i.e.~$\mathfrak{m}' = \mathcal{T}(\mathfrak{m})$, and the matrix of $\mathcal{T}$ in the vectorised picture is $T = M_{\rm in}M^{\prime -1}_{\rm in} = M^{-1}_{\rm out}M'_{\rm out}$. For any initial state $\rho$ in $\mathfrak{m}$ and the corresponding initial state $\rho'$ in $\mathfrak{m}'$, we always have $M_{\rm out}\Dket{\rho} = M'_{\rm out}\Dket{\rho'}$ because of the distribution equivalence. Then, $\Dket{\rho'} = T^{-1} \Dket{\rho}$. For any measurement operator $E$ in $\mathfrak{m}$ and the corresponding measurement operator $E'$ in $\mathfrak{m}'$, we always have $\Dbra{E}M_{\rm in} = \Dbra{E'}M'_{\rm in}$ because of the distribution equivalence. Then, $\Dbra{E'} = \Dket{E} T$. For any map $\mathcal{M}$ in $\mathfrak{m}$ and the corresponding map $\mathcal{M}'$ in $\mathfrak{m}'$, we always have $M_{\rm out}MM_{\rm in} = M'_{\rm out}M'M'_{\rm in}$ because of the distribution equivalence. Then, $M' = T^{-1}MT$.

\begin{lemma}
If complete sets of initial states and measurement operators are related by a transformation, two representations are related by the same transformation.
\label{the:GST}
\end{lemma}

Suppose complete sets of initial states and measurement operators are related by the transformation $\mathcal{S}$, i.e.~$\rho'_i = \mathcal{S}^{-1}(\rho_i)$ and $E'_k = \mathcal{S}^*(E_k)$, where $i,k = 1,\ldots,d_{\rm H}^2$, two representations are related by the same transformation $\mathcal{S}$. $S$ is the matrix of $\mathcal{S}$ in the vectorised picture, then $M'_{\rm in} = S^{-1}M_{\rm in}$ and $T = S$.

\section{Theorem 2: sufficient condition of uniqueness}
\label{sec:Theorem2}

According to Lemma~\ref{the:completeness}, $\mathfrak{m}'$ and $\mathfrak{m}$ are related by a similarity transformation on maps, i.e.~$\mathfrak{m}' = \mathcal{T}(\mathfrak{m})$, because of the completeness of $\Pi_{\rm QPT}$.

\subsection{Basis states}

First, we consider the probability
\begin{eqnarray}
\Tr[\mathcal{T}^*(\pi_a) \mathcal{T}^{-1}(\pi_a)] = 1.
\end{eqnarray}
Here, $\mathcal{T}^{-1}(\pi_a)$ and $\mathcal{T}^*(\pi_a)$ are physical state and measurement operator, respectively. Without loss of generality, we suppose that the state $\mathcal{T}^{-1}(\pi_a)$ is a diagonal matrix, and $N_a$ is the orthogonal projection onto the null space of $\mathcal{T}^{-1}(\pi_a)$. Because the probability is $1$, diagonal elements of $\mathcal{T}^*(\pi_a)$ are $1$ in the subspace of $(\openone-N_a)$, and all off-diagonal elements related to the subspace of $(\openone-N_a)$ are $0$. Here, we have used that $\mathcal{T}^*(\pi_a) \leq \openone$. Then, we have
\begin{eqnarray}
(\openone-N_a)\mathcal{T}^*(\pi_a)(\openone-N_a) = (\openone-N_a)
\end{eqnarray}
and
\begin{eqnarray}
(\openone-N_a)\mathcal{T}^*(\pi_a)N_a = N_a\mathcal{T}^*(\pi_a)(\openone-N_a) = 0.
\end{eqnarray}
Therefore,
\begin{eqnarray}
\mathcal{T}^*(\pi_a) = (\openone-N_a) + N_a \mathcal{T}^*(\pi_a) N_a.
\end{eqnarray}

Then, we consider the probability
\begin{eqnarray}
\Tr[\mathcal{T}^*(\pi_b) \mathcal{T}^{-1}(\pi_a)] = 0,
\end{eqnarray}
where $b\neq a$. Without loss of generality, we suppose that the state $\mathcal{T}^{-1}(\pi_a)$ is a diagonal matrix. Because the probability is $0$, diagonal elements of $\mathcal{T}^*(\pi_b)$ are $0$ in the subspace of $(\openone-N_a)$, and all off-diagonal elements related to the subspace of $(\openone-N_a)$ are $0$. Here, we have used that $\mathcal{T}^*(\pi_b) \geq 0$. Then, we have
\begin{eqnarray}
(\openone-N_a)\mathcal{T}^*(\pi_b)(\openone-N_a) = 0.
\end{eqnarray}
Therefore,
\begin{eqnarray}
(\openone-N_a)(\openone-N_b)(\openone-N_a) = 0
\label{eq:orthogonality}
\end{eqnarray}
and
\begin{eqnarray}
(\openone-N_a)N_b \mathcal{T}^*(\pi_b) N_b(\openone-N_a) = 0.
\end{eqnarray}
Here, we have used that $\openone-N_b$ and $N_b \mathcal{T}^*(\pi_b) N_b$ are positive semidefinite.

The dimension of the Hilbert space is $d_{\rm H}$, and there are $d_{\rm H}$ orthogonal subspaces $\{ \openone-N_a \}$ [see Eq.~(\ref{eq:orthogonality})]. Therefore, $\{\openone-N_a\}$ are one-dimensional subspaces, and $N_a \mathcal{T}^*(\pi_a) N_a = 0$. We use $\ket{a'}$ to denote the state of the subspace $\openone-N_a$, then $\{ \ket{a'} \}$ is an orthonormal basis of the Hilbert space, and
\begin{eqnarray}
\mathcal{T}^*(\pi_a) = \mathcal{T}^{-1}(\pi_a) = \ketbra{a'}{a'}.
\end{eqnarray}

\subsection{Pure states and projective measurements}

Similar to $\pi_a$, for all $\pi \in \Pi$, because
\begin{eqnarray}
\Tr[\mathcal{T}^*(\pi) \mathcal{T}^{-1}(\pi)] = 1.
\end{eqnarray}
we have
\begin{eqnarray}
\mathcal{T}^*(\pi) = (\openone-N) + N \mathcal{T}^*(\pi) N,
\end{eqnarray}
where $N$ is the orthogonal projection onto the null space of $\mathcal{T}^{-1}(\pi)$. Because
\begin{eqnarray}
\sum_a \Tr[\mathcal{T}^*(\pi) \mathcal{T}^{-1}(\pi_a)] = 1,
\end{eqnarray}
we have $\Tr[\mathcal{T}^*(\pi)] = 1$. Then, $\openone-N$ is a one-dimensional subspace, and $N \mathcal{T}^*(\pi) N = 0$. Therefore, $\mathcal{T}^*(\pi) = \mathcal{T}^{-1}(\pi)$ are one-dimensional orthogonal projections, i.e.~all $\{ \mathcal{T}^{-1}(\pi) \}$ are pure states, and all $\{ \mathcal{T}^*(\pi) \}$ are projective measurements.

\subsection{Two-basis states and measurement operators}

We consider probabilities
\begin{eqnarray}
&& \Tr[\mathcal{T}^*(\pi_a) \mathcal{T}^{-1}(\pi^{\rm x}_{a,b})] \notag \\
&=& \Tr[\mathcal{T}^*(\pi_b) \mathcal{T}^{-1}(\pi^{\rm x}_{a,b})] = \frac{1}{2}.
\end{eqnarray}
According to the probability distribution, the pure state of $\mathcal{T}^{-1}(\pi^{\rm x}_{a,b})$ can be written as
\begin{eqnarray}
\ket{\psi^{\rm x\prime}_{a,b}} = \frac{1}{\sqrt{2}}(\ket{a'} + e^{i\phi^{\rm x}_{a,b}}\ket{b'}),
\end{eqnarray}
and
\begin{eqnarray}
\mathcal{T}^*(\pi^{\rm x}_{a,b}) = \mathcal{T}^{-1}(\pi^{\rm x}_{a,b}) = \ketbra{\psi^{\rm x\prime}_{a,b}}{\psi^{\rm x\prime}_{a,b}}.
\end{eqnarray}

Similarly,
\begin{eqnarray}
\mathcal{T}^*(\pi^{\rm y}_{a,b}) = \mathcal{T}^{-1}(\pi^{\rm y}_{a,b}) = \ketbra{\psi^{\rm y\prime}_{a,b}}{\psi^{\rm y\prime}_{a,b}},
\end{eqnarray}
where
\begin{eqnarray}
\ket{\psi^{\rm y\prime}_{a,b}} = \frac{1}{\sqrt{2}}(\ket{a'} + e^{i\phi^{\rm y}_{a,b}}\ket{b'}).
\end{eqnarray}

Because of the probability
\begin{eqnarray}
\Tr[\mathcal{T}^*(\pi^{\rm x}_{a,b}) \mathcal{T}^{-1}(\pi^{\rm y}_{a,b})] = \frac{1}{2},
\end{eqnarray}
we have
\begin{eqnarray}
e^{i\pi^{\rm y}_{a,b}} = i \kappa_{a,b} e^{i\phi^{\rm x}_{a,b}},
\label{eq:xyab}
\end{eqnarray}
where $\kappa_{a,b} = \pm 1$.

\subsection{Three-basis states and measurement operators}

First, we consider probabilities
\begin{eqnarray}
&& \Tr[\mathcal{T}^*(\pi_a) \mathcal{T}^{-1}(\pi^{\rm x}_{a,b,c})] \notag \\
&=& \Tr[\mathcal{T}^*(\pi_b) \mathcal{T}^{-1}(\pi^{\rm x}_{a,b,c})] \notag \\
&=& \Tr[\mathcal{T}^*(\pi_c) \mathcal{T}^{-1}(\pi^{\rm x}_{a,b,c})] = \frac{1}{3}.
\end{eqnarray}
According to the probability distribution, the pure state of $\mathcal{T}^{-1}(\pi^{\rm x}_{a,b,c})$ can be written as
\begin{eqnarray}
\ket{\psi^{\rm x\prime}_{a,b,c}} = \frac{1}{\sqrt{3}}(\ket{a'} + e^{i\mu^{\rm x}_{a,b,c}} \ket{b'} + e^{i\nu^{\rm x}_{a,b,c}} \ket{c'}),
\end{eqnarray}
and
\begin{eqnarray}
\mathcal{T}^*(\pi^{\rm x}_{a,b,c}) = \mathcal{T}^{-1}(\pi^{\rm x}_{a,b,c}) = \ketbra{\psi^{\rm x\prime}_{a,b,c}}{\psi^{\rm x\prime}_{a,b,c}}.
\end{eqnarray}

Then, we consider probabilities
\begin{eqnarray}
&& \Tr[\mathcal{T}^*(\pi^{\rm x}_{a,b}) \mathcal{T}^{-1}(\pi^{\rm x}_{a,b,c})] \notag \\
&& \Tr[\mathcal{T}^*(\pi^{\rm x}_{a,c}) \mathcal{T}^{-1}(\pi^{\rm x}_{a,b,c})] \notag \\
&=& \Tr[\mathcal{T}^*(\pi^{\rm x}_{b,c}) \mathcal{T}^{-1}(\pi^{\rm x}_{a,b,c})] = \frac{2}{3}.
\end{eqnarray}
According to these probabilities, we have
\begin{eqnarray}
e^{i\mu^{\rm x}_{a,b,c}} &=& e^{i\phi^{\rm x}_{a,b}}, \label{eq:mux} \\
e^{i\nu^{\rm x}_{a,b,c}} &=& e^{i\phi^{\rm x}_{a,c}} \label{eq:nux}
\end{eqnarray}
and
\begin{eqnarray}
e^{i\phi^{\rm x}_{b,c}} = e^{i(\nu^{\rm x}_{a,b,c} - \mu^{\rm x}_{a,b,c})} = e^{i(\phi^{\rm x}_{a,c} - \phi^{\rm x}_{a,b})}.
\label{eq:xabc}
\end{eqnarray}

Similarly,
\begin{eqnarray}
\mathcal{T}^*(\pi^{\rm y}_{a,b,c}) = \mathcal{T}^{-1}(\pi^{\rm y}_{a,b,c}) = \ketbra{\psi^{\rm y\prime}_{a,b,c}}{\psi^{\rm y\prime}_{a,b,c}},
\end{eqnarray}
where
\begin{eqnarray}
\ket{\psi^{\rm y\prime}_{a,b,c}} = \frac{1}{\sqrt{3}}(\ket{a'} + e^{i\mu^{\rm y}_{a,b,c}} \ket{b'} + e^{i\nu^{\rm y}_{a,b,c}} \ket{c'}),
\end{eqnarray}
\begin{eqnarray}
e^{i\mu^{\rm y}_{a,b,c}} &=& e^{i\phi^{\rm y}_{a,b}}, \label{eq:muya} \\
e^{i\nu^{\rm y}_{a,b,c}} &=& e^{i\phi^{\rm y}_{a,c}} \label{eq:nuya}
\end{eqnarray}
and
\begin{eqnarray}
e^{i\phi^{\rm x}_{b,c}} = e^{i(\nu^{\rm y}_{a,b,c} - \mu^{\rm y}_{a,b,c})} = e^{i(\phi^{\rm y}_{a,c} - \phi^{\rm y}_{a,b})}.
\label{eq:yabc}
\end{eqnarray}
We would like to remark that it is $e^{i\phi^{\rm x}_{b,c}}$ rather than $e^{i\phi^{\rm y}_{b,c}}$ in the above equation.

We also have
\begin{eqnarray}
\mathcal{T}^*(\pi^{\rm y}_{c,b,a}) = \mathcal{T}^{-1}(\pi^{\rm y}_{c,b,a}) = \ketbra{\psi^{\rm y\prime}_{c,b,a}}{\psi^{\rm y\prime}_{c,b,a}},
\end{eqnarray}
where
\begin{eqnarray}
\ket{\psi^{\rm y\prime}_{c,b,a}} = \frac{1}{\sqrt{3}}(\ket{c'} + e^{i\mu^{\rm y}_{c,b,a}} \ket{b'} + e^{i\nu^{\rm y}_{c,b,a}} \ket{a'}),
\end{eqnarray}
\begin{eqnarray}
e^{i\mu^{\rm y}_{c,b,a}} &=& -e^{-i\phi^{\rm y}_{b,c}}, \label{eq:muyc} \\
e^{i\nu^{\rm y}_{c,b,a}} &=& -e^{-i\phi^{\rm y}_{a,c}} \label{eq:nuyc}
\end{eqnarray}
and
\begin{eqnarray}
e^{i\phi^{\rm x}_{a,b}} = e^{i(\mu^{\rm y}_{c,b,a} - \nu^{\rm y}_{c,b,a})} = e^{i(\phi^{\rm y}_{a,c}-\phi^{\rm y}_{b,c})}.
\label{eq:ycba}
\end{eqnarray}

\subsection{Transformation}

According to Eq.~(\ref{eq:xabc}), we have
\begin{eqnarray}
e^{i\phi^{\rm x}_{a,b}} = e^{i(\phi^{\rm x}_{1,b} - \phi^{\rm x}_{1,a})}
\end{eqnarray}
for all $a<b$, where $\phi^{\rm x}_{1,1} = 1$. According to Eq.~(\ref{eq:mux}) and Eq.~(\ref{eq:nux}), we have
\begin{eqnarray}
e^{i\mu^{\rm x}_{a,b,c}} &=& e^{i(\phi^{\rm x}_{1,b} - \phi^{\rm x}_{1,a})}, \\
e^{i\nu^{\rm x}_{a,b,c}} &=& e^{i(\phi^{\rm x}_{1,c} - \phi^{\rm x}_{1,a})}.
\end{eqnarray}

According to Eq.~(\ref{eq:xyab}), Eq.~(\ref{eq:xabc}) and Eq.~(\ref{eq:yabc}), we have
\begin{eqnarray}
e^{i(\phi^{\rm y}_{a,c} - \phi^{\rm y}_{a,b})} = \kappa_{a,c}\kappa_{a,b}e^{i(\phi^{\rm x}_{a,c} - \phi^{\rm x}_{a,b})} = e^{i(\phi^{\rm x}_{a,c} - \phi^{\rm x}_{a,b})}.~~~~
\end{eqnarray}
i.e.~$\kappa_{a,c} = \kappa_{a,b}$. Similarly, according to Eq.~(\ref{eq:xyab}), Eq.~(\ref{eq:xabc}) and Eq.~(\ref{eq:ycba}), we have
\begin{eqnarray}
e^{i(\phi^{\rm y}_{a,c}-\phi^{\rm y}_{b,c})} = \kappa_{a,c}\kappa_{b,c}e^{i(\phi^{\rm x}_{a,c} - \phi^{\rm x}_{b,c})} = e^{i(\phi^{\rm x}_{a,c} - \phi^{\rm x}_{b,c})}.~~~~
\end{eqnarray}
i.e.~$\kappa_{a,c} = \kappa_{b,c}$. Then, $\kappa_{a,b} = \kappa_{1,2}$ for all $a<b$. Therefore,
\begin{eqnarray}
e^{i\phi^{\rm y}_{a,b}} &=& i\kappa_{1,2}e^{i(\phi^{\rm x}_{1,b} - \phi^{\rm x}_{1,a})}, \\
e^{i\mu^{\rm y}_{a,b,c}} &=& i\kappa_{1,2}e^{i(\phi^{\rm x}_{1,b} - \phi^{\rm x}_{1,a})}, \\
e^{i\nu^{\rm y}_{a,b,c}} &=& i\kappa_{1,2}e^{i(\phi^{\rm x}_{1,c} - \phi^{\rm x}_{1,a})}, \\
e^{i\mu^{\rm y}_{c,b,a}} &=& i\kappa_{1,2}e^{i(\phi^{\rm x}_{1,b} - \phi^{\rm x}_{1,c})}, \\
e^{i\nu^{\rm y}_{c,b,a}} &=& i\kappa_{1,2}e^{i(\phi^{\rm x}_{1,a} - \phi^{\rm x}_{1,c})}.
\end{eqnarray}
Here, we have used Eq.~(\ref{eq:muya}), Eq.~(\ref{eq:nuya}), Eq.~(\ref{eq:muyc}) and Eq.~(\ref{eq:nuyc}).

Therefore, projections in two model representations $\mathfrak{m}$ and $\mathcal{T}(\mathfrak{m})$ are related by the transformation $\mathcal{S} = \mathcal{W}^{(1-\kappa_{1,2})/2}\mathcal{U}^{-1}$, i.e.~
\begin{eqnarray}
\mathcal{S}^*(\pi) = \mathcal{S}^{-1}(\pi) = \mathcal{T}^*(\pi) = \mathcal{T}^{-1}(\pi)
\end{eqnarray}
for all $\pi\in\Pi$, where $\mathcal{U}(\bullet) = u\bullet u^\dag$,
\begin{eqnarray}
u = \sum_a e^{i\phi^{\rm x}_{1,a}} \ketbra{a'}{a}
\end{eqnarray}
and $\mathcal{S}^* = \mathcal{S}^{-1} = \mathcal{U}\mathcal{W}^{(1-\kappa_{1,2})/2}$. If $\kappa_{1,2} = +1$, $\mathcal{S}$ is a unitary transformation. If $\kappa_{1,2} = -1$, $\mathcal{S}$ is an anti-unitary transformation.

Because projections $\Pi_{\rm QPT}$ are complete, two representations are related by the transformation $\mathcal{S}$, i.e.~$\mathcal{S}(\mathfrak{m}) = \mathcal{T}(\mathfrak{m})$, according to Lemma~\ref{the:GST}. The transformation $\mathcal{S}$ is either unitary or anti-unitary, therefore $\mathfrak{m}$ and $\mathcal{T}(\mathfrak{m})$ are always representations of the same model. Theorem~2 is proved.

\section{Lemma 2: symmetry transformations on unitary maps}
\label{sec:Lemma2}

In Ref.~[22] in the main text, it is proved that a physical map is unitary if and only if its determinant is $1$. The similarity transformation preserves the determinant of a map. Therefore, for two physical maps $\mathcal{U}$ and $\mathcal{U}' = \mathcal{T}^{-1}\mathcal{U}\mathcal{T}$, if $\mathcal{U}$ is unitary, $\mathcal{U}'$ is also unitary. Are such two maps related by a unitary or anti-unitary transformation? It is obvious when $d_{\rm H} = 2,3$, and we conjecture that it is true for all dimensions. Later we will prove that it is true for super-non-degenerate unitary maps.

If eigenvalues of the unitary matrix $u$ are $\{ e^{i\theta_a} \}$, eigenvalues of the map $\mathcal{U}$ are $\{ e^{i(\theta_a-\theta_b)} \}$. The map is super-non-degenerate if $e^{i(\theta_c-\theta_d)}e^{i(\theta_e-\theta_f)} \notin \{ e^{i(\theta_a-\theta_b)} \}$ for all $a,b,c,d$ satisfying $c\neq d$, $e\neq f$, $c\neq f$ and $d\neq e$.

\begin{lemma}
When $d_{\rm H} = 2,3$, $\mathcal{U},\mathcal{T}^{-1}\mathcal{U}\mathcal{T} \in \mathbb{U}$ if and only if there exists $\mathcal{S}  \in \mathbb{S}$ such that $\mathcal{T}^{-1}\mathcal{U}\mathcal{T} = \mathcal{S}^{-1}\mathcal{U}\mathcal{S}$.
\label{the:dim2and3}
\end{lemma}

\begin{figure}[tbp]
\centering
\includegraphics[width=1\linewidth]{\figpath 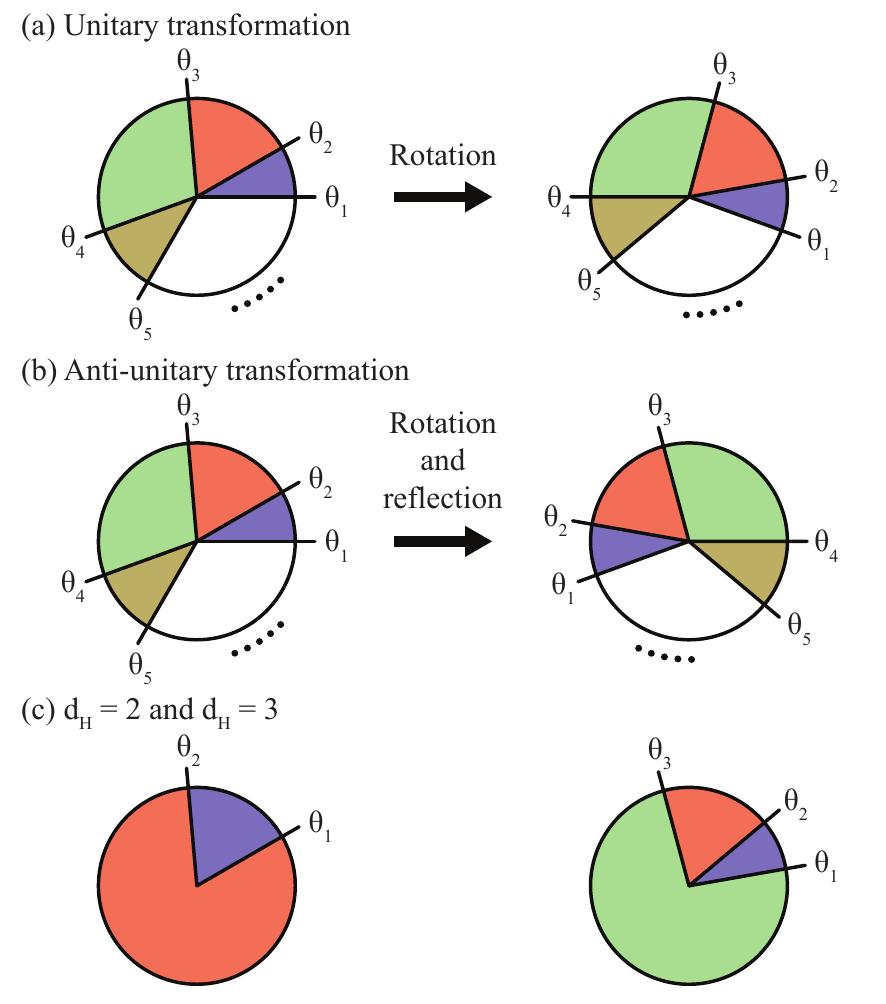}
\caption{
Each eigenvalue of a unitary matrix corresponds to a line on the circle as shown in the figure. (a) If two circles are related by a rotation, then corresponding unitary maps are related by a unitary transformation. (b) If two circles are related by a rotation and a reflection, then corresponding unitary maps are related by an anti-unitary transformation. (c) When $d_{\rm H} = 2,3$, there is only one configuration of lines (up to the rotation and reflection) given the size of each sector.
}
\label{fig:pizza}
\end{figure}

If $\mathcal{U}(\bullet) = u\bullet u^\dag$ and eigenvalues of $u$ are $\{ e^{i\theta_a} \}$, then eigenvalues of $\mathcal{U}$ are $\{ e^{i(\theta_a-\theta_b)} \}$. Similarly, if $\mathcal{U}'(\bullet) = \mathcal{T}^{-1}\mathcal{U}\mathcal{T}(\bullet) = u'\bullet u^{\prime\dag}$ and eigenvalues of $u'$ are $\{ e^{i\theta'_a} \}$, then eigenvalues of $\mathcal{U}'$ are $\{ e^{i(\theta'_a-\theta'_b)} \}$. If $e^{i\theta_a} = e^{i\omega}e^{i\theta'_a}$ for all $a$, $\mathcal{U}$ and $\mathcal{U}'$ are related by a unitary transformation, see Fig.~\ref{fig:pizza}(a). If $e^{i\theta_a} = e^{i\omega}e^{-i\theta'_a}$ for all $a$, $\mathcal{U}$ and $\mathcal{U}'$ are related by an anti-unitary transformation, see Fig.~\ref{fig:pizza}(b).

If $d_{\rm H} = 2,3$, each pair of $e^{i(\theta_a-\theta_b)}$ and $e^{i(\theta_b-\theta_a)}$ with $a\neq b$ corresponds to the size of a sector in Fig.~\ref{fig:pizza}(c). Therefore, $\mathcal{U}$ and $\mathcal{U}'$ are always related by a unitary or anti-unitary transformation. Lemma~\ref{the:dim2and3} is proved.

\begin{lemma}
Let $\mathcal{U} \in \mathbb{U}$ be super-non-degenerate. Then $\mathcal{T}^{-1}\mathcal{U}\mathcal{T} \in \mathbb{U}$ if and only if there exists $\mathcal{S}  \in \mathbb{S}$ such that $\mathcal{T}^{-1}\mathcal{U}\mathcal{T} = \mathcal{S}^{-1}\mathcal{U}\mathcal{S}$.
\label{the:SND}
\end{lemma}

Two sets of eigenvalues are the same because of the similarity transformation, i.e.~$E = \{ e^{i(\theta_a-\theta_b)} \} = \{ e^{i(\theta'_a-\theta'_b)} \}$. Therefore $\lambda^*\in E$ for all eigenvalues $\lambda\in E$.

Without loss of generality, we suppose that $e^{i(\theta_1-\theta_2)} = e^{i(\theta'_1-\theta'_2)}$. Then $E_{1,2} = \{ e^{i(\theta_1-\theta_a)}, e^{i(\theta_b-\theta_2)} ~\vert~ a,b\neq 1,2 \} = \{ e^{i(\theta'_1-\theta'_a)}, e^{i(\theta'_b-\theta'_2)} ~\vert~ a,b\neq 1,2 \}$ are all eigenvalues $\lambda \in E$ satisfying $\lambda^* e^{i(\theta_1-\theta_2)} \in E - \{1\}$, because of the super-non-degeneracy.

Either $e^{i(\theta_1-\theta_3)} = e^{i(\theta'_1-\theta'_a)}$ or $e^{i(\theta_1-\theta_3)} = e^{i(\theta'_b-\theta'_2)}$. Without loss of generality, we suppose that $e^{i(\theta_1-\theta_3)} = e^{i(\theta'_1-\theta'_3)}$ or $e^{i(\theta_1-\theta_3)} = e^{i(\theta'_3-\theta'_2)}$.

If $e^{i(\theta_1-\theta_3)} = e^{i(\theta'_1-\theta'_3)}$, $E_{1,2,3} = \{ e^{i(\theta_1-\theta_a)} ~\vert~ a\neq 1,2,3 \} = \{ e^{i(\theta'_1-\theta'_a)} ~\vert~ a\neq 1,2,3 \}$ are all eigenvalues $\lambda \in E_{1,2}$ satisfying $\lambda^* e^{i(\theta_1-\theta_3)} \in E - \{1\}$. Without loss of generality, we suppose that $e^{i(\theta_1-\theta_a)} = e^{i(\theta'_1-\theta'_a)}$, where $a\neq 1,2,3$. Then, $e^{i\theta_a} = e^{i(\theta_1-\theta'_1)}e^{i\theta'_a}$ for all $a$. Therefore, $\mathcal{U}$ and $\mathcal{U}'$ are related by a unitary transformation.

If $e^{i(\theta_1-\theta_3)} = e^{i(\theta'_3-\theta'_2)}$, $E_{1,2,3} = \{ e^{i(\theta_1-\theta_a)} ~\vert~ a\neq 1,2,3 \} = \{ e^{i(\theta'_b-\theta'_2)} ~\vert~ b\neq 1,2,3 \}$ are all eigenvalues $\lambda \in E_{1,2}$ satisfying $\lambda^* e^{i(\theta_1-\theta_3)} \in E - \{1\}$. Without loss of generality, we suppose that $e^{i(\theta_1-\theta_a)} = e^{i(\theta'_a-\theta'_2)}$, where $a\neq 1,2,3$. Then, $e^{i\theta_1} = e^{i(\theta_1+\theta'_2)}e^{-i\theta'_2}$, $e^{i\theta_2} = e^{i(\theta_1+\theta'_2)}e^{-i\theta'_1}$ and $e^{i\theta_a} = e^{i(\theta_1+\theta'_2)}e^{-i\theta'_a}$ for all $a \neq 1,2$. Therefore, $\mathcal{U}$ and $\mathcal{U}'$ are related by an anti-unitary transformation.

Lemma~\ref{the:SND} is proved.

\begin{lemma}
The set of super-non-degenerate unitary maps is dense.
\label{the:dense}
\end{lemma}

Without loss of generality, we consider a diagonal unitary matrix $u = {\rm diag}(e^{i\theta_1},\ldots,e^{i\theta_{d_{\rm H}}})$. We define $u^{(N)} = {\rm diag}(e^{i\theta^{(N)}_1},\ldots,e^{i\theta^{(N)}_{d_{\rm H}}})$, where
\begin{eqnarray}
\frac{\theta^{(N)}_a}{2\pi} = \left\lfloor 10^{N}\frac{\theta_a}{2\pi} \right\rfloor 10^{-N} + 2^{-a}10^{-3N}.
\end{eqnarray}
One can check that $u^{(N)}$ is super-non-degenerate. Let $\mathcal{U}^{(N)}$ be the map of $u^{(N)}$, then
\begin{eqnarray}
\lim_{N\rightarrow \infty} \norm{ \mathcal{U}^{(N)}-\mathcal{U} } = 0.
\end{eqnarray}
Lemma~\ref{the:dense} is proved.

Suppose that $\mathcal{T}^{-1}\mathcal{U}^{(N)}\mathcal{T} = \mathcal{S}^{(N)-1}\mathcal{U}^{(N)}\mathcal{S}^{(N)}$, where $\norm{\bullet}$ is a multiplicative norm, then we have
\begin{eqnarray}
&& \norm{ \mathcal{T}^{-1}\mathcal{U}^{(N)}\mathcal{T} - \mathcal{T}^{-1}\mathcal{U}\mathcal{T} } \notag \\
&\leq & \norm{\mathcal{T}^{-1}}\norm{\mathcal{T}}\norm{ \mathcal{U}^{(N)}-\mathcal{U} }, \\
&& \norm{ \mathcal{S}^{(N)-1}\mathcal{U}^{(N)}\mathcal{S}^{(N)} - \mathcal{S}^{(N)-1}\mathcal{U}\mathcal{S}^{(N)} } \notag \\
&\leq & \norm{\mathcal{S}^{(N)-1}}\norm{\mathcal{S}^{(N)}}\norm{ \mathcal{U}^{(N)}-\mathcal{U} }.
\end{eqnarray}
Therefore,
\begin{eqnarray}
&& \norm{ \mathcal{S}^{(N)-1}\mathcal{U}\mathcal{S}^{(N)} - \mathcal{T}^{-1}\mathcal{U}\mathcal{T} } \notag \\
&\leq & (\norm{\mathcal{T}^{-1}}\norm{\mathcal{T}}+\norm{\mathcal{S}^{(N)-1}}\norm{\mathcal{S}^{(N)}})\norm{ \mathcal{U}^{(N)}-\mathcal{U} }
\end{eqnarray}
and
\begin{eqnarray}
\lim_{N\rightarrow \infty} \norm{ \mathcal{S}^{(N)-1}\mathcal{U}\mathcal{S}^{(N)} - \mathcal{T}^{-1}\mathcal{U}\mathcal{T} } = 0.
\end{eqnarray}
Taking $\mathcal{S}_\mathcal{U} = \lim_{N\rightarrow \infty} \mathcal{S}^{(N)}$, we have $\mathcal{T}^{-1}\mathcal{U}\mathcal{T} = \mathcal{S}_\mathcal{U}^{-1}\mathcal{U}\mathcal{S}_\mathcal{U}$.

\begin{lemma}
Let $\mathcal{T}$ such that, for all $\mathcal{U} \in \mathbb{U}$, there exists $\mathcal{S}_\mathcal{U} \in \mathbb{S}$ with $\mathcal{T}^{-1}\mathcal{U}\mathcal{T} = \mathcal{S}_\mathcal{U}^{-1}\mathcal{U}\mathcal{S}_\mathcal{U}$. Here, $\mathcal{S}_\mathcal{U}$ depends on $\mathcal{U}$. Then, there exists an $\mathcal{S} \in \mathbb{S}$ such that $\mathcal{T}^{-1}\mathcal{U}\mathcal{T} = \mathcal{S}^{-1}\mathcal{U}\mathcal{S}$ for all $\mathcal{U} \in \mathbb{U}$.
\label{the:group}
\end{lemma}

% Suppose $\mathcal{U}(\bullet) = u\bullet u^\dag$, we have $\mathcal{U}'(\bullet) = \mathcal{S}_\mathcal{U}^{-1}\mathcal{U}\mathcal{S}_\mathcal{U}(\bullet) = u'\bullet u^{\prime\dag}$, where $u' = \mathcal{S}_\mathcal{U}^{-1}(u)$ if $\mathcal{S}_\mathcal{U}$ is unitary and $u' = \mathcal{S}_\mathcal{U}^{-1}(u^\dag)$ if $\mathcal{S}_\mathcal{U}$ is anti-unitary. When $d_{\rm H} = 2$, $u^\dag = \mathcal{W}\mathcal{Y}(u)$ for all $u$, where $\mathcal{Y}(\bullet) = Y\bullet Y$, and $Y$ is the Pauli operator. Then, $\mathcal{U}' = \mathcal{S}_\mathcal{U}^{\prime-1}\mathcal{U}\mathcal{S}_\mathcal{U}'$ when $\mathcal{S}_\mathcal{U}$ is anti-unitary, where $\mathcal{S}_\mathcal{U}' = \mathcal{Y}\mathcal{W}\mathcal{S}_\mathcal{U}$ is unitary. Therefore, when $d_{\rm H} = 2$, for each $\mathcal{U} \in \mathbb{U}$, there exists $\mathcal{S}_\mathcal{U} \in \mathbb{U}$ such that $\mathcal{T}^{-1}\mathcal{U}\mathcal{T} = \mathcal{S}_\mathcal{U}^{-1}\mathcal{U}\mathcal{S}_\mathcal{U}$. In the following, we only consider unitary transformations when $d_{\rm H} > 2$.

We only need to consider generators of the unitary group $\{ u_{\psi,\theta} \}$, where
\begin{eqnarray}
u_{\psi,\theta} = e^{i\ketbra{\psi}{\psi}\theta} = \openone + (e^{i\theta}-1)\ketbra{\psi}{\psi},
\end{eqnarray}
and corresponding maps are $\{ \mathcal{U}_{\psi,\theta}(\bullet) = u_{\psi,\theta} \bullet u_{\psi,\theta}^\dag \}$. Using the transformation $\mathcal{S}_\psi = \mathcal{S}_{\mathcal{U}_{\psi,\pi/2}}$, we have
\begin{eqnarray}
\mathcal{T}^{-1}\mathcal{U}_{\psi,\pi/2}\mathcal{T} = \mathcal{S}_\psi^{-1}\mathcal{U}_{\psi,\pi/2}\mathcal{S}_\psi = \mathcal{U}_{s^{-1}(\psi),\kappa_\psi \pi/2},
\end{eqnarray}
where $\ketbra{s^{-1}(\psi)}{s^{-1}(\psi)} = \mathcal{S}_\psi^{-1}(\ketbra{\psi}{\psi})$, $\kappa_\psi = +1$ if $\mathcal{S}_\psi$ is unitary, and $\kappa_\psi = -1$ if $\mathcal{S}_\psi$ is anti-unitary. Here, we have defined a transformation on the ray space, i.e.
\begin{eqnarray}
s^{-1}(\ketbra{\psi}{\psi}) = \mathcal{S}_\psi^{-1}(\ketbra{\psi}{\psi}).
\end{eqnarray}
% When $d_{\rm H} = 2$, we can always take $\kappa_\psi = +1$.

Using $\mathcal{U}_{\psi,\pi/2}^n$, where $n = 0,1,2$, we can expand $\mathcal{U}_{\psi,\theta}$ as
\begin{eqnarray}
\mathcal{U}_{\psi,\theta} &=& \frac{\mathcal{U}_{\psi,\pi/2}^0 + \mathcal{U}_{\psi,\pi/2}^2}{2} \notag \\
&&+ \cos\theta \frac{\mathcal{U}_{\psi,\pi/2}^0 - \mathcal{U}_{\psi,\pi/2}^2}{2} \notag \\
&&+ \sin\theta \left( \mathcal{U}_{\psi,\pi/2}^1 - \frac{\mathcal{U}_{\psi,\pi/2}^0 + \mathcal{U}_{\psi,\pi/2}^2}{2} \right).
\end{eqnarray}
Then using the linearity of the similarity transformation and $\mathcal{T}^{-1}\mathcal{U}_{\psi,\pi/2}^n\mathcal{T} = \mathcal{U}_{s^{-1}(\psi),\kappa_\psi \pi/2}^n$, we have
\begin{eqnarray}
\mathcal{T}^{-1} \mathcal{U}_{\psi,\theta} \mathcal{T} = \mathcal{U}_{s^{-1}(\psi),\kappa_\psi \theta}.
\end{eqnarray}

We consider two arbitrary states $\ket{\psi}$ and $\ket{\phi}$. Without loss of generality, we assume that
\begin{eqnarray}
\ket{\psi} &=& \cos\frac{\alpha}{2}\ket{0} + \sin\frac{\alpha}{2}\ket{1}, \\
\ket{\phi} &=& \cos\frac{\alpha}{2}\ket{0} - \sin\frac{\alpha}{2}\ket{1}.
\end{eqnarray}
Then $\abs{\braket{\psi}{\phi}} = \abs{\cos\alpha}$ and
\begin{eqnarray}
\ketbra{\psi}{\psi} &=& \frac{1}{2}(I + \cos\alpha Z + \sin\alpha X ), \\
\ketbra{\phi}{\phi} &=& \frac{1}{2}(I + \cos\alpha Z - \sin\alpha X ),
\end{eqnarray}
where $I,X,Y,Z$ are Pauli operators in the subspace of $\{ \ket{0},\ket{1} \}$. Corresponding unitary matrices can be expressed as
\begin{eqnarray}
u_{\psi,\theta} &=& (\openone-I) + e^{i\frac{\theta}{2}} \cos\frac{\theta}{2} I \notag \\
&&+ i e^{i\frac{\theta}{2}} \sin\frac{\theta}{2} (\cos\alpha Z + \sin\alpha X) \\
u_{\phi,\omega} &=& (\openone-I) + e^{i\frac{\omega}{2}} \cos\frac{\omega}{2} I \notag \\
&&+ i e^{i\frac{\omega}{2}} \sin\frac{\omega}{2} (\cos\alpha Z - \sin\alpha X).
\end{eqnarray}

Because of the similarity transformation, two maps $\mathcal{U}_{\psi,\theta}\mathcal{U}_{\phi,\omega}$ and $\mathcal{U}_{s^{-1}(\psi),\kappa_\psi \theta}\mathcal{U}_{s^{-1}(\phi),\kappa_\phi \omega} = \mathcal{T}^{-1} \mathcal{U}_{\psi,\theta}\mathcal{U}_{\phi,\omega} \mathcal{T}$ have the same spectrum for all $\theta$ and $\omega$. The unitary matrix of $\mathcal{U}_{\psi,\theta}\mathcal{U}_{\phi,\omega}$ is
\begin{eqnarray}
u_{\psi,\theta} u_{\phi,\omega} = (\openone-I) + e^{i\frac{\theta+\omega}{2}} (\cos\delta I + i\sin\delta A),
\end{eqnarray}
where
\begin{eqnarray}
\cos\delta &=& \cos\frac{\theta+\omega}{2}\cos^2\alpha + \cos\frac{\theta-\omega}{2}\sin^2\alpha, \\
\sin\delta A &=& \sin\frac{\theta+\omega}{2}\cos\alpha Z + \sin\frac{\theta-\omega}{2}\sin\alpha X \notag \\
&&+ (\cos\frac{\theta-\omega}{2} - \cos\frac{\theta+\omega}{2})\frac{\sin2\alpha}{2} Y
\end{eqnarray}
and $A^2 = I$. If $\abs{\omega},\abs{\theta} \in [0,\pi)$, we have $\abs{\frac{\theta+\omega}{2}},\abs{\frac{\theta-\omega}{2}} \in [0,\pi)$, and $\abs{\delta}$ is between $\abs{\frac{\theta+\omega}{2}}$ and $\abs{\frac{\theta-\omega}{2}}$. Therefore, two maps have the same spectrum for all $\theta$ and $\omega$ if and only if $\kappa_\psi = \kappa_\phi$ and $\abs{\braket{\psi}{\phi}} = \abs{\braket{s^{-1}(\psi)}{s^{-1}(\phi)}}$.

According to Theorem~\ref{the:Wigner}, we have
\begin{eqnarray}
s^{-1}(\ketbra{\psi}{\psi}) = \mathcal{S}^{-1}(\ketbra{\psi}{\psi}),
\end{eqnarray}
where $\mathcal{S}^{-1} \in \mathbb{S}$. Then,
\begin{eqnarray}
\mathcal{S}^{-1}\mathcal{U}_{\psi,\pi}\mathcal{S} &=& \mathcal{U}_{s^{-1}(\psi),\pm\pi} \notag \\
&=& \mathcal{U}_{s^{-1}(\psi),\kappa_\psi\pi} = \mathcal{T}^{-1}\mathcal{U}_{\psi,\pi}\mathcal{T}.
\end{eqnarray}
Because
\begin{eqnarray}
u_{\psi,\pi} u_{\phi,\pi} = (\openone-I) + (\cos2\alpha I - i\sin2\alpha Y),
\end{eqnarray}
$\{ \mathcal{U}_{\psi,\pi} \}$ is the set of generators of $\mathbb{U}$. Therefore
\begin{eqnarray}
\mathcal{S}^{-1}\mathcal{U}\mathcal{S} = \mathcal{T}^{-1}\mathcal{U}\mathcal{T}
\end{eqnarray}
for all $\mathcal{U} \in \mathbb{U}$. Lemma~\ref{the:group} is proved.

It is obvious that, $\mathcal{T}^{-1}\mathcal{U}\mathcal{T} \in \mathbb{U}$ for all $\mathcal{U} \in \mathbb{U}$ if there exists a unitary or anti-unitary transformation $\mathcal{S}\in \mathbb{S}$ such that $\mathcal{S}^{-1}\mathcal{U}\mathcal{S} = \mathcal{T}^{-1}\mathcal{U}\mathcal{T}$ for all $\mathcal{U} \in \mathbb{U}$, i.e.~the `if' statement of Lemma 2 in the main text is true. If $\mathcal{T}^{-1}\mathcal{U}\mathcal{T} \in \mathbb{U}$ for all $\mathcal{U} \in \mathbb{U}$, there exists $\mathcal{S}_\mathcal{U} \in \mathbb{S}$ (which depends on $\mathcal{U}$) with $\mathcal{T}^{-1}\mathcal{U}\mathcal{T} = \mathcal{S}_\mathcal{U}^{-1}\mathcal{U}\mathcal{S}_\mathcal{U}$ for all $\mathcal{U} \in \mathbb{U}$, because super-non-degenerate unitary maps are dense (Lemma~\ref{the:dense}). Then, according to Lemma~\ref{the:group}, there exists an $\mathcal{S}  \in \mathbb{S}$ such that $\mathcal{T}^{-1}\mathcal{U}\mathcal{T} = \mathcal{S}^{-1}\mathcal{U}\mathcal{S}$ for all $\mathcal{U} \in \mathbb{U}$. The `only if' statement of Lemma 2 in the main text is proved.

\section{Theorem 3: sufficient condition of uniqueness}
\label{sec:Theorem3}

\begin{lemma}
Let $A$ be a non-trivial Hermitian matrix, i.e.~$A \neq \Tr(A)d_{\rm H}^{-1}\openone$, then $\{ \mathcal{U}(A) \vert \mathcal{U}\in \mathbb{U} \}$ is complete.
\label{the:completeness_generation}
\end{lemma}

Without loss of generality, we suppose that $A = \sum_a A_a \ketbra{a}{a}$ is a diagonal matrix with diagonal elements in the descending order $A_1 \geq A_2 \geq \ldots \geq A_{d_{\rm H}}$. Because $A$ is non-trivial, $A_1 \neq A_{d_{\rm H}}$.

We consider unitary matrices
\begin{eqnarray}
u_{b,c} = \ketbra{b}{c} + \ketbra{c}{b} + \sum_{a\neq b,c} \ketbra{a}{a}
\end{eqnarray}
and
\begin{eqnarray}
u_{\rm loop} = \ketbra{1}{1} + \ketbra{d_{\rm H}}{2} + \sum_{a=3}^{d_{\rm H}} \ketbra{a-1}{a}.
\end{eqnarray}
Corresponding maps are $\mathcal{U}_{b,c}(\bullet) = u_{b,c}\bullet u_{b,c}^\dag$ and $\mathcal{U}_{\rm loop}(\bullet) = u_{\rm loop}\bullet u_{\rm loop}^\dag$.

Using these maps, we have
\begin{eqnarray}
A' &=& \frac{1}{d_{\rm H}-1} \sum_{l=1}^{d_{\rm H}-1} \mathcal{U}_{\rm loop}^l (A) \notag \\
&=& A_1\ketbra{1}{1} + \frac{\Tr(A)-A_1}{d_{\rm H}-1} \sum_{a=2}^{d_{\rm H}} \ketbra{a}{a}
\end{eqnarray}
and
\begin{eqnarray}
A'' &=& \frac{1}{d_{\rm H}-1} \sum_{l=1}^{d_{\rm H}-1} \mathcal{U}_{\rm loop}^l \mathcal{U}_{1,d_{\rm H}} (A) \notag \\
&=& A_{d_{\rm H}}\ketbra{1}{1} + \frac{\Tr(A)-A_{d_{\rm H}}}{d_{\rm H}-1} \sum_{a=2}^{d_{\rm H}} \ketbra{a}{a}.
\end{eqnarray}
Then,
\begin{eqnarray}
\ketbra{1}{1} = \frac{[\Tr(A)-A_{d_{\rm H}}]A' - [\Tr(A)-A_1]A''}{[\Tr(A)-A_{d_{\rm H}}]A_1 - [\Tr(A)-A_1]A_{d_{\rm H}}}
\end{eqnarray}
is a linear combination of matrices in $\{ \mathcal{U}(A) \vert \mathcal{U}\in \mathbb{U} \}$. Because $A_1 \neq A_{d_{\rm H}}$, the denominator $[\Tr(A)-A_{d_{\rm H}}]A_1 - [\Tr(A)-A_1]A_{d_{\rm H}} = (A_1-A_{d_{\rm H}})\Tr(A) \neq 0$.

Because $\{ \mathcal{U}(\ketbra{1}{1}) \vert \mathcal{U}\in \mathbb{U} \}$ is complete, $\{ \mathcal{U}(A) \vert \mathcal{U}\in \mathbb{U} \}$ is also complete. Lemma~\ref{the:completeness_generation} is proved.

Suppose $\rho \in \{\rho_i\}$ and $E \in \{E_k\}$ are singular, then both of them are non-trivial. Therefore, $\{ \mathcal{U}(\rho) \vert \mathcal{U}\in \mathbb{U} \}$ and $\{ \mathcal{U}(E) \vert \mathcal{U}\in \mathbb{U} \}$ are complete. Then, according to Lemma~\ref{the:completeness}, $\mathfrak{m}'$ and $\mathfrak{m}$ are related by a similarity transformation on maps, i.e.~$\mathfrak{m}' = \mathcal{T}(\mathfrak{m})$.

\begin{lemma}
A map $\mathcal{D}$ commutes with all unitary maps, i.e.~$[\mathcal{D}, \mathcal{U}] = 0$ for all $\mathcal{U}\in \mathbb{U}$, if and only if $\mathcal{D}$ is a depolarising map.
\label{the:depolarising}
\end{lemma}

We consider the state $\ketbra{0}{0}$ and unitary matrices $u_{0,\theta}$ and
\begin{eqnarray}
v = \ketbra{0}{0} + v',
\end{eqnarray}
where $v' = (\openone - \ketbra{0}{0})v'(\openone - \ketbra{0}{0})$ can be any unitary matrix in the subspace of $\openone - \ketbra{0}{0}$. Then, we have
\begin{eqnarray}
\rho = \mathcal{D}(\ketbra{0}{0}) = \mathcal{D}(u_{0,\theta}\ketbra{0}{0}u_{0,\theta}^\dag) = u_{0,\theta} \rho u_{0,\theta}^\dag
\end{eqnarray}
for all $\theta$. Therefore,
\begin{eqnarray}
\rho = f\ketbra{0}{0} + \rho',
\end{eqnarray}
where $f = \bra{0}\rho\ket{0}$ and $\rho' = (\openone-\ketbra{0}{0}) \rho (\openone-\ketbra{0}{0})$. We also have
\begin{eqnarray}
\rho = v \rho v^\dag = f\ketbra{0}{0} + v'\rho' v^{\prime\dag}.
\end{eqnarray}
Because $\rho' = v'\rho' v^{\prime\dag}$ for all $v'$,
\begin{eqnarray}
\rho' = \frac{1-f}{d_{\rm H}-1}(\openone - \ketbra{0}{0}).
\end{eqnarray}
Therefore,
\begin{eqnarray}
\rho = (f-\frac{1-f}{d_{\rm H}-1})\ketbra{0}{0} + \frac{1-f}{d_{\rm H}-1}\openone.
\end{eqnarray}

For any state $\ketbra{\psi}{\psi} = u\ketbra{0}{0}u^\dag$, we have
\begin{eqnarray}
\mathcal{D}(\ketbra{\psi}{\psi}) = u\rho u^\dag = F\ketbra{\psi}{\psi} + (1-F)d_{\rm H}^{-1}\openone,
\end{eqnarray}
where $F = f-(1-f)/(d_{\rm H}-1)$. Therefore, $\mathcal{D} = \mathcal{D}_F$ is a depolarising map.

\end{document}